\documentclass{emulateapj}
\usepackage{epsfig}

\newcommand{\saxj}{\mbox{SAX J1808.4$-$3658}}
\newcommand{\RXTE}{\textit{RXTE}}
\newcommand{\us}{$\mu$s}
\newcommand{\uHz}{$\mu$Hz}
\newcommand{\fluxunits}{~erg~cm$^{-2}$~s$^{-1}$}

\begin{document}

\title{A decade of timing an accretion-powered millisecond pulsar:\\
  The continuing spin down and orbital evolution of \saxj}
\shorttitle{A Decade of Timing \saxj}
\shortauthors{Hartman et~al.}

\submitted{Accepted by ApJ}

\author{
  Jacob M. Hartman\altaffilmark{1,2},
  Alessandro Patruno\altaffilmark{3},
  Deepto Chakrabarty\altaffilmark{4},
  Craig B. Markwardt\altaffilmark{5,6},
  Edward H. Morgan\altaffilmark{4},
  Michiel van der Klis\altaffilmark{3},
  Rudy Wijnands\altaffilmark{3} }
\altaffiltext{1}{Space Science Division, Naval Research Laboratory, 
  Washington, DC 20375, USA; jacob.hartman@nrl.navy.mil}
\altaffiltext{2}{National Research Council Research Associate}
\altaffiltext{3}{Astronomical Institute ``Anton Pannekoek,''
  University of Amsterdam, Kruislaan 403, 1098 SJ Amsterdam, Netherlands;
  a.patruno, m.b.m.vanderklis, r.a.d.wijnands@uva.nl}
\altaffiltext{4}{Department of Physics and Kavli Institute for Astrophysics
  and Space Research, Massachusetts Institute of Technology, Cambridge, MA
  02139; deepto, ehm@space.mit.edu}
\altaffiltext{5}{CRESST/Department of Astronomy, University of Maryland, 
  College Park, MD 20742}
\altaffiltext{6}{Astrophysics Science Division, NASA Goddard Space Flight 
  Center, Greenbelt, MD 20771; craigm@milkyway.gsfc.nasa.gov}

\begin{abstract}

The {\em Rossi X-ray Timing Explorer} has observed five outbursts from the
transient 2.5~ms accretion-powered pulsar \saxj\ during 1998--2008.  We
present a pulse timing study of the most recent outburst and compare it with
the previous timing solutions.  The spin frequency of the source continues to
decrease at a rate of $(-5.5\pm1.2)\times10^{-18}$~Hz~s$^{-1}$, which is
consistent with the previously determined spin derivative.  The spin-down
occurs mostly during quiescence, and it is most likely due to the magnetic
dipole torque from a $B = 1.5\times10^8$~G dipolar field at the neutron star
surface.  We also find that the 2~hr binary orbital period is increasing at a
rate of $(3.80\pm0.06)\times10^{-12}$~s~s$^{-1}$, also consistent with
previous measurements.  It remains uncertain whether this orbital change
reflects secular evolution or short-term variability.

\end{abstract}



\section{Introduction}

Ten years ago, observations of the recurrent X-ray transient \saxj\ with the
{\em Rossi X-ray Timing Explorer} (\RXTE) revealed coherent 2.5~ms pulsations,
establishing this source as the first known accretion-powered millisecond
pulsar (AMP; \citealt{Wijnands98}).  Its immediate recognition as the
``missing link'' in which old pulsars are spun up to millisecond periods by
accretion in an X-ray binary was only the beginning of what can be learned
from this class of sources: among other areas of theoretical interest, AMPs
continue to reveal insights into the nature of magnetically channeled
accretion flow, the torques that act upon neutron stars in low-mass X-ray
binaries (LMXBs), and the orbital evolution of these systems.  This paper
investigates the 2008 October X-ray outburst of \saxj, which touches upon all
these questions.

\saxj\ was originally discovered by the {\em BeppoSAX} WFC monitor during a
1996 outburst \citep{IntZand98, IntZand01}.  It was the first LMXB known to
exhibit all three modes of millisecond X-ray variability established by the
\RXTE: accretion-powered pulsations \citep{Wijnands98}, X-ray burst
oscillations \citep{Chakrabarty03}, and kilohertz quasi-periodic oscillations
\citep{Wijnands03}.  The source has been extensively studied for over 1.7 Ms
with the \RXTE\ PCA during outbursts in 1998, 2000, 2002, 2005, and 2008,
making it the best-studied AMP.  Its outbursts rise from quiescence ($L_{\rm
bol} \lesssim 10^{32}$~erg~s$^{-1}$; \citealt{Campana02, Heinke07}) to their
peak fluxes in $\approx$5~d, followed by $\approx$5~d at peak bolometric
luminosities of $L_{\rm bol} = ($5--8$)\times10^{36}$~erg~s$^{-1}$ (for a
distance of 3.5~kpc and the bolometric correction given by
\citealt{Galloway06}), $\approx$15~d of decay back to near-quiescent fluxes,
and a month or more of low-luminosity flaring (\citealt{Hartman08}, hereafter
H08).  These outbursts recur every 1.6--3.3~years.

Orbital modulation of the persistent pulsations establishes a 2.01~hr binary
orbital period \citep{Chakrabarty98}.  Its mass donor is likely an extremely
low-mass ($\approx$0.05~$M_\sun$) brown dwarf \citep{Bildsten01}.  The system
has been detected and extensively studied in the optical \citep{Roche98,
Giles99, Wang01, Deloye08, Wang08}, and its relatively high optical luminosity
during X-ray quiescence has led to speculation that the neutron star (NS) may
be an active radio pulsar during these intervals \citep{Homer01, Burderi03,
Campana04}, although radio pulsations have not been detected \citep{Burgay03}.
Pulse timing of the four outbursts during 1998--2005 revealed that the spin
frequency of \saxj\ is decreasing during quiescence at a rate of $\dot{\nu} =
(-5.6\pm2.0)\times10^{-16}$~Hz~s$^{-1}$, while its orbital period is
increasing at a rate of $\dot{P}_{\rm orb} =
(3.5\pm0.2)\times10^{-12}$~s~s$^{-1}$ (H08; \citealt{DiSalvo08}).

\section{X-ray observations and analysis}

Routine monitoring of the Galactic bulge region with the \RXTE\ PCA revealed
\saxj\ to again be in outburst on 2008 Sep 21 \citep{Markwardt08}.  \RXTE\
observed the source almost every day during the outburst, producing 270~ks of
data with 122~\us\ time resolution and 64 energy channels covering the
2--60~keV range of the PCA.  Unfortunately, the total effective area of the
PCA has decreased significantly since the earlier outbursts: an average of
only 1.8 of the 5 proportional counter units are on during these observations,
resulting in a mean effective area of 2200~cm$^2$.

Our analysis followed the techniques developed in H08 and applied to the
earlier outbursts of \saxj.  We shifted the photon arrival times to the solar
system barycenter using the optical position from H08, applied the \RXTE\ fine
clock correction, and filtered out data during Earth occultations and
intervals of unstable pointing.  We also removed data from 1~min before to
10~min after the two thermonuclear X-ray bursts observed by the \RXTE\ during
this outburst.

To measure the times of arrival (TOAs) and fractional amplitudes of the
persistent pulsations, we folded 512~s intervals of data using a preliminary
timing model based on extrapolating forward the H08 results.  The timing
models were applied and fitted using the TEMPO pulse timing program, version
11.005,\footnote{At http://wwww.atnf.csiro.au/research/pulsar/tempo} and
assumed a circular orbit and a fixed spin frequency.  From the resulting
folded pulse profiles, we measured the phases and amplitudes of the
fundamental and the second harmonic of the pulsation.\footnote{Throughout this
paper, we number the harmonics such that the $k$th harmonic is $k$ times the
frequency of the 401~Hz fundamental.}  Higher harmonics had insufficient power
to be useful for pulse phase timing.

Special care must be taken when measuring the pulse TOAs for \saxj.  As in
past outbursts, the harmonic phase residuals relative to our best timing
models exhibit variability on time scales of $\sim$10~hr and longer that is
well in excess of the Poisson noise expected from counting statistics.  As
described in H08, we applied a frequency-domain filter that weighs the
harmonics according to observed noise properties in order to generate a
minimum-variance estimator of the NS spin phase.  For this outburst, the
second harmonic had less excess timing noise, so it was weighted more heavily
when estimating the overall pulse TOAs.  With these pulse TOA estimates, we
used TEMPO to fit an improved timing model.  By iteratively re-folding the
data, estimating the pulse TOAs, and fitting a new model, we converged on our
best-fit timing model.

Uncertainties in our frequency measurements were estimated using Monte Carlo
simulations, following the prescription of H08.  These simulations account for
the impact of long--time-scale correlations in the pulse arrival times.  To
test for the presence of a frequency derivative, we fit one to the pulse TOAs
and estimated its significance with our simulations.

To measure the orbital period and its derivative, we connected the orbital
phases of all the outbursts.  The uncertainty in the 2008 orbital phase was
estimated by measuring and accounting for the timing noise at scales of
$\lesssim P_{\rm orb}$; as in previous outbursts (H08), the noise was close to
white but somewhat higher than what would be expected form Poisson noise
alone.

\section{Results}

\begin{figure}[t]
  \includegraphics[width=0.465\textwidth]{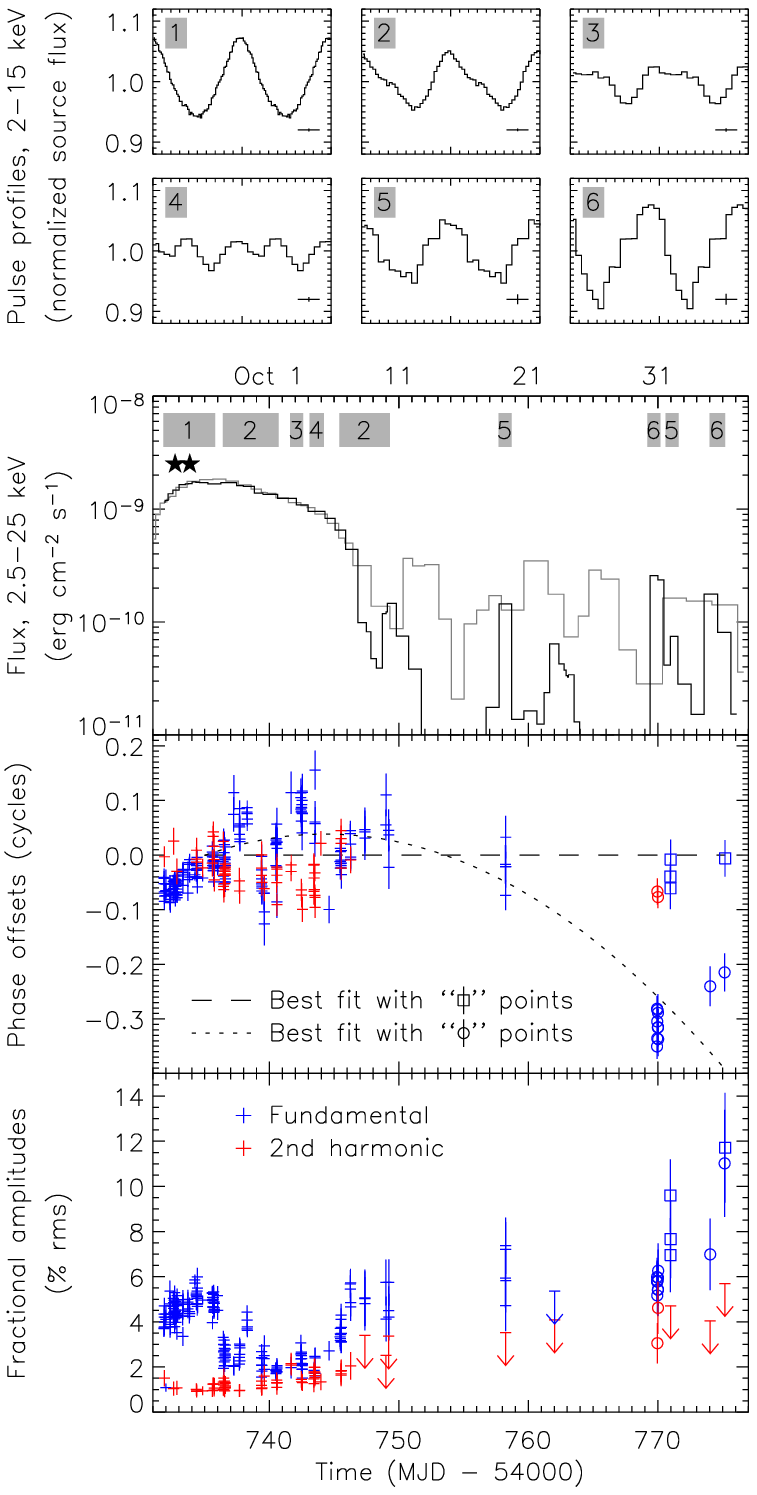}
  \caption{The pulse profiles, light curve, phase residuals, and fractional
    amplitudes for the 2008 outburst of \saxj.  The top panels show the
    2--15~keV pulse profiles observed during the outburst.  The profiles are
    background-subtracted and normalized such that the phase bins have a mean
    value of unity, and the error crosses in the lower right of each graph
    give the mean uncertainties.  Each box shows two cycles, so each minor
    tick mark on the phase axis is 0.1 cycles.  The phases are based on our
    best-fit constant-frequency spin model.  The grey boxes in the light curve
    plot indicate the time intervals during which the profiles were observed.
    The second panel shows the light curve of the 2008 outburst, in black.
    The remarkably similar 2005 outburst is shown in grey for comparison.  The
    stars indicate the times of thermonuclear bursts during 2008.  The third
    panel shows the pulse phase residuals relative to our best-fit
    constant-frequency model, $\nu = 400.975210133$~Hz.  Blue points indicate
    the phase of the fundamental; red points indicate the second harmonic.
    Each point represents 512~s of timing data.  The phase jumps after MJD
    54760 led us to separate the phase measurements as earlier (circles) and
    later (squares), and the dashed and dotted lines show the respective phase
    models for each group.  The bottom plot shows the fractional amplitudes of
    the fundamental and harmonic.  Arrows give 90\% upper limits. \bigskip}
  \label{fig:Outburst}
\end{figure}

In most respects, the 2008 outburst of \saxj\ continues the trends and
patterns observed during the earlier outbursts.  Figure~\ref{fig:Outburst}
shows its light curve, pulse phases, and fractional amplitudes, along with six
representative 2--15~keV pulse profiles.

The light curve of the main body of the outburst is remarkably similar the
light curve observed in 2005, which is included in Figure~\ref{fig:Outburst}
for comparison.  The 2008 outburst is slightly dimmer (maximum 2.5--25~keV
flux of $(1.74\pm0.02)\times10^{-9}$\fluxunits) and its light curve has two
maxima, but the outburst rise, the length of the peak, and the slow decay down
to $\approx$$4\times10^{-10}$\fluxunits are nearly identical.

The flaring tail of the new outburst is markedly dimmer.  During the two
previous outbursts for which the tail was observed immediately following the
main outburst (2002 and 2005), the flares were quasiperiodic with a 4--6~d
time scale, and the 2.5--25~keV flux varied between
$\sim$$10^{-11}$\fluxunits\ and $4\times10^{-10}$\fluxunits.  The source
remained consistently detectable for at least a month following the main
outburst.  In contrast, the brightest observations during the tail of the 2008
outburst only reach $2.5\times10^{-10}$\fluxunits, and during Oct 13--17, the
source was not detectable at all with the \RXTE\ ($\sim$$10^{-11}$\fluxunits\
sensitivity).  This behavior is similar to the intermittent detections during
the tail of the 2000 outburst \citep{Wijnands01}.

Most of the pulse profiles were similar to what we observed during the earlier
outbursts: profiles 1--3 of Figure~\ref{fig:Outburst} closely resemble
profiles 1--3 in Figure~3 of H08.  The evolution of the pulse shape progresses
in the same way.  As seen in the 2005 and (to a lesser extent) 2002 outbursts,
a symmetric pulse during the burst rise becomes increasingly skewed during the
burst peak.  In the phase residuals, this pulse shape change manifests as a
progressively increasing phase of the fundamental, while the phase of the
second harmonic remains constant; this behavior can be seen during MJD
54732--54735 in the phase offset plot of Figure~\ref{fig:Outburst}.  Since
these pulse shape changes confuse measurement of the spin phase, we omit the
outburst rise and early peak (points before MJD 54735) from our timing models.
This choice is admittedly somewhat arbitrary, as the pulse profile continues
to change throughout the outburst, albeit more slowly.  However, this same
region was omitted from the analysis of the earlier outbursts in H08, so we
remove it here so that our frequency measurements are directly
comparable.\footnote{Including the phases during the outburst rise decreases
the frequencies cited throughout this paper by 20--30~nHz.  Including this
stage in the earlier outbursts causes their measured frequencies to shift
similarly, so the value of the long-term spin down is not affected.}

The clear double peaks of profile~4 are an exception to these similarities in
the pulse shape evolution.  This profile folds 8.2~ks of data during MJD
54743.2--54744.0.  It occurs during the slow-decay stage of the outburst, at a
2.5--25~keV flux of $1.0\times10^{-9}$\fluxunits.  In none of the earlier
outbursts was a second peak so prominent, and the weak secondary peaks that
were observed occured exclusively during the flaring tail, when the source was
a factor of 2--3 dimmer.

The most unusual timing feature of this outburst is the presence of abrupt
shifts of the fundamental's phase during the outburst tail.  These phase jumps
are clear at MJD 54770 and 54774 in the phase offset plot of
Figure~\ref{fig:Outburst}, at which times the fundamental phases are 0.2--0.3
cycles earlier than predicted by the constant-frequency model shown.  The
observations with earlier phases were folded and summed to produce profile~6,
and their phases are marked by circles.  The rest of the observations in the
tail were folded to produce profile~5, and after MJD 54760 their phases are
marked by squares.  While substantial phase shifts have been observed in the
previous outbursts of \saxj\ (compare Fig.~2 of \citealt{Burderi06} and Fig.~1
of H08), these jumps were particularly large and rapid.  During the 23~hours
between the observations on MJD 54770 and 54771, the phase shifted by
0.26~cycles; during the single observation on MJD 54775, the phase shifted by
0.20~cycles in 40~minutes.  These shifts are due in part to differing pulse
shapes.  The centroid of the pulse --- to which the fundamental is most
sensitive --- arrives appreciably earlier in profile~6, but the peaks of the
two profiles are separated by a smaller difference.  However, this nuance is
lost unless many observations are folded together (as in the displayed
profiles), and for most observations in the tail only the fundamental is
available for pulse timing.

Given this limitation, the inclusion of these phase measurements of the
fundamental in our timing models is subject to interpretation.  If these
points are all excluded, the phase residuals for MJD 54735--54760 give a
frequency of 400.97521012(3)~Hz.  (Parenthetical digits give $1\;\sigma$
estimates of the uncertainties.)  If the circled points are used and the
square points excluded, the timing solution requires a large apparent spin up
to account for the advancing phases in the outburst tail.  Its frequency rises
at a rate $\dot{\nu} = (1.1\pm0.2)\times10^{-13}$~Hz~s$^{-1}$ from a value of
$\nu_{\rm start} = 400.97521002(5)$ at the start of the outburst.  This fit is
indicated by the dotted line in the phase offset panel of
Figure~\ref{fig:Outburst}.  It does not completely eliminate the jumps: a
substantial difference (0.16~cycles) remains between the points around MJD
54770 and 54774.  On the other hand, if we only include the square points in
the outburst tail, the data are consistent with a constant spin frequency of
$400.97521013(2)$~Hz.  This selection eliminates all phase jumps greater than
0.1~cycles.  Finally, if we include all the points in the tail, we get
$\dot{\nu} = 0.6(4)\times10^{-13}$~Hz~s$^{-1}$ and $\nu_{\rm start} =
400.97521009(6)$, although the phase jumps result in a poor fit and large
uncertainties.

\begin{figure}[t]
  \includegraphics[width=0.47\textwidth]{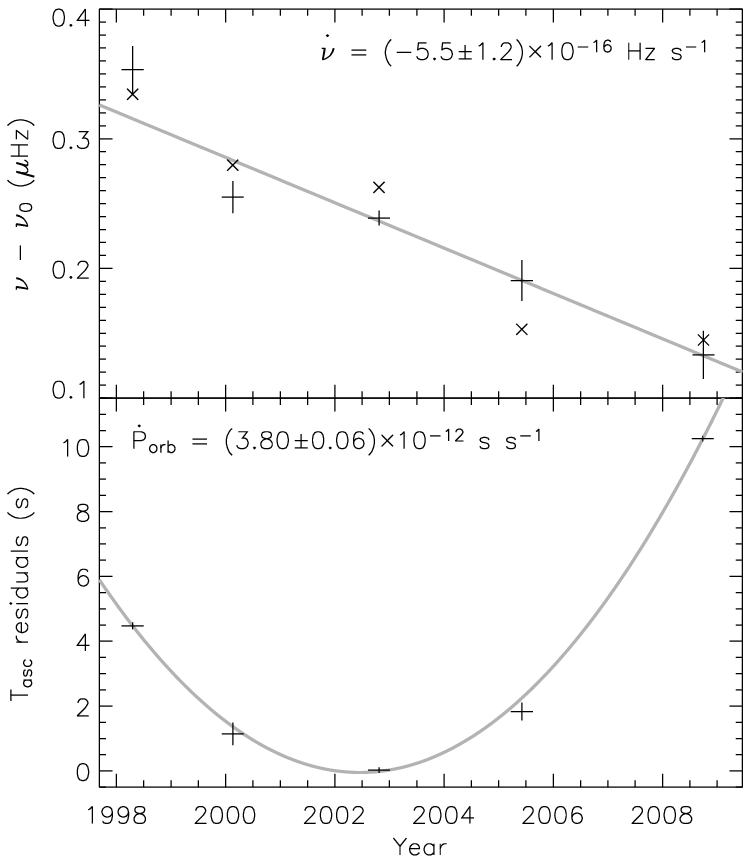}
  \caption{The changing spin and orbital timing of \saxj\ during the
    1998--2008 outbursts.  Top: constant-frequency measurements of each
    outburst, showing the spin down of the star.  The frequencies are relative
    to $\nu_0 = 400.97521000$~Hz.  The $\times$'s mark what the frequencies
    would be if the fit source position differed from the actual position by
    $2\,\sigma$ along the ecliptic plane in the direction of increasing RA.
    Bottom: the observed times of ascending node, relative to the expected
    times for a constant period.  The $T_{\rm asc}$ of each outburst comes
    progressively later, indicating a positive orbital period derivative.
    \bigskip}
  \label{fig:Timing}
\end{figure}

Ultimately we are most interested in these frequency measurements to determine
whether the long-term spin down observed in H08 has continued, and on this
point the data are clear: regardless of our assumptions, the estimated spin
frequency at the start of the 2008 outburst is less than the 2005 frequency of
400.97521019(2)~Hz (H08).  Fitting a constant slope through the frequencies
observed for the five outbursts yields a long-term spin-down rate of
$\dot{\nu} = (-5.5\pm1.2)\times10^{-16}$~Hz~s$^{-1}$, as shown in the top
panel of Figure~\ref{fig:Timing}.  This rate and figure use the 2008 timing
model obtained when we only include the square points; we prefer this model on
physical grounds (see \S\ref{sect:DiscPhaseJumps}).  That said, using the
$\nu_{\rm start}$ frequency from the model for only the circled points does
not greatly alter the result, slightly increasing its magnitude to
$-6.2\times10^{-16}$~Hz~s$^{-1}$.  However, the 2008 point becomes an outlier
and the $\chi^2$ of the fit increases.

These values are in excellent agreement with the spin down of
$(-5.6\pm2.0)\times10^{-16}$~Hz~s$^{-1}$ found for the 1998--2005 outbursts
(H08).  The frequency uncertainties estimated using our Monte Carlo
simulations give $\chi^2 = 9.7$ with 3~degrees of freedom.  To account for
this large $\chi^2$, we scaled up the frequency uncertainties such that the
reduced $\chi^2$ is unity when calculating the uncertainty for $\dot{\nu}$.
The fit is not particularly sensitive to the source position:  shifting it by
$1\;\sigma$ along the ecliptic (using the uncertainty from the optical
position in H08) changes the spin down by $0.2\times10^{-16}$~Hz~s$^{-1}$.  No
position significantly decreases the $\chi^2$ of the linear trendline.

The farthest outlier from a constant frequency derivative is the 2000
outburst, of which we only observed the flaring tail.  As noted in H08, this
frequency measurement may be complicated by small systematic differences in
the observed pulse frequency between the main outburst and its tail, which
were observed during the 1998 and 2002 outbursts.  Removing this outburst
lowers the $\chi^2$ to 1.9 (2 degrees of freedom), consistent with a constant
derivative, and changes the long-term spin down to
$(-6.5\pm0.8)\times10^{-16}$~Hz~s$^{-1}$.

The measured time of ascending node\footnote{The ascending node is the point
of zero degrees true longitude, which is the same as zero degrees mean
longitude for a circular orbit like this one.} continues to advance relative
to a constant orbital period.  A quadratic provides a good fit ($\chi^2 = 2.9$
with 2~degrees of freedom), consistent with a constant orbital period
derivative of $\dot{P}_{\rm orb} = (3.80\pm0.06)\times10^{-12}$~s~s$^{-1}$.
This measurement is $1.5\;\sigma$ greater than the value in H08.  It is in
good agreement with \citet{Burderi09}, which measured the orbital phase of the
2008 outburst using {\em XMM-Newton} data.  All other orbital parameters
differ by less than $1\;\sigma$ from the values listed in H08.
Table~\ref{tbl:ParamSummary} summarizes all the parameters for the long-term
pulse timing of \saxj.

Proper interpretation of the phase residuals hinges on an accurate handling of
the orbital parameters, particularly the orbital period, due to the sparsity
of data in the outburst's tail.  From the orbital parameters derived by
connecting the advancing orbital phases of the 1998--2005 outbursts (H08), we
extrapolate a 2008 orbital period of 7249.15763(6)~s.  Including the 2008
outburst's phase, this method gives 7249.15771(2)~s.  Fitting the orbital
parameters using the fundamental for the first half of the ouburst only
(cf.~\citealt{Patruno08}) or using only the second harmonic give compatible
periods of 7249.160(3)~s and 7249.153(5)~s, respectively.  However, fitting to
the phase of the fundamental over the entire outburst yields a period of
7249.174(2)~s, $9\;\sigma$ higher than the far more accurate values derived
through phase connection.  Because the phase measurements in the outburst tail
are sparse due to the source's highly variable flux, their coverage of the
orbital phase is limited, and TEMPO attempts to ``correct'' for the observed
jumps by over-fitting $P_{\rm orb}$.

\begin{deluxetable}{@{} lr @{}}
\tabletypesize{\footnotesize}
\tablecolumns{2}
\tablewidth{0pt}
\tablecaption{Combined timing parameters for \saxj
  \label{tbl:ParamSummary}}
\startdata
\hline\hline\\[-1.5 ex]
Orbital period, $P_{\rm orb}$ (s) \tablenotemark{a} &
  7249.156980(4)\\
Orbital period derivative, $\dot{P}_{\rm orb}$ ($10^{-12}$~s~s$^{-1}$) &
  3.80(6)\\
Projected semimajor axis, $a_{\rm x} \sin i$ (light-ms) &
  62.812(2)\\
Time of ascending node, $T_{\rm asc}$ (MJD, TDB) &
  52499.9602472(9)\\
Eccentricity, $e$ (95\% confidence upper limit) &
  $< 1.2\times 10^{-4}$\\
Spin frequency, $\nu$ (Hz) \tablenotemark{a} &
  400.975210240(9)\\
Spin frequency derivative, $\dot\nu$ ($10^{-16}$~Hz~s$^{-1}$) &
  $-$5.5(1.2)\\[-2 ex]
\enddata
\tablenotetext{a}{$P_{\rm orb}$ and $\nu$ are specified for the time
  $T_{\rm asc}$ (2002 Aug 14.0).}
\end{deluxetable}

\section{Discussion}

\subsection{The double-peaked pulse profiles and phase jumps}
\label{sect:DiscPhaseJumps}

The presence of double-peaked pulse profiles during the 2008 outburst was
notable both because two peaks were so well defined (compare with the weaker
secondary peaks shown in Fig.~3 of H08) and because they occurred while the
outburst was still relatively bright: 2--3 times more luminous than when
double-peaked profiles were first seen during the 2002 and 2005 outbursts.
H08 speculated that the appearance of a second peak may be caused by the
recession of the accretion disk as the accretion rate drops, revealing the
star's antipodal hot spot.  \citet{Ibragimov09} went further, estimating from
magnetic accretion theory and geometric considerations that the appearance of
the antipodal hot spot at fluxes of $F_{\rm bol} \approx
0.8\times10^{-9}$\fluxunits\ during 2002 and 2005 would imply a surface
magnetic field of $B\approx(0.4$--$1.2)\times10^8$~G.

The detection of double-peaked profiles at $F_{\rm bol} =
2.1\times10^{-9}$\fluxunits\ during 2008 requires this field estimate to be
increased somewhat, although its weak dependence on flux ($\propto F^{1/2}$ in
eq.~14 of \citealt{Ibragimov09}) means that is still not in conflict with the
$B \leq 1.5\times10^8$~G limit implied by the long-term spin down (H08; see
also \S\ref{sect:DiscSpinDown}).  More troublesome is the question of why the
second spot does not remain visible as the accretion rate continues to drop.
It is difficult to imagine a change in the geometry or inner radius of the
disk that temporarily reveals the antipodal hot spot but has no impact at all
on the accretion rate and flux, as is the case here.  More exotic
explanations, such as a temporary bifurcation of the accretion column to
produce two hot spots, may be needed; simulations have suggested that such
configurations are easily possible if higher-order field components are
present \citep{Long08}.

The large phase jumps in the tail of the 2008 outburst are the most unusual
feature of the pulse profiles, with the phase of the fundamental in one case
jumping by 0.20~cycles in only 40~minutes.  Clearly these are not changes in
the rotational phase of the NS, which complicates their inclusion while
constructing a timing solution.  We considered two extremes: that the earlier
points reflected the NS spin phase, which necessitated a spin up of
$1.1(2)\times10^{-13}$~Hz~s$^{-1}$; and that the later points tracked the
spin, in which case a constant-frequency timing solution provides a good fit.

We prefer the latter case on physical grounds.  From standard magnetic
accretion torque theory (e.g., \citealt{Pringle72}), an accretion rate of
$\dot{M}$ can impart a maximum torque of $2\pi I \dot{\nu} \leq \dot{M} (G M
r_{\rm co})^{1/2}$, where $r_{\rm co}$ is the co-rotation radius of the NS and
$I$ is its moment of inertia.  Estimating $\dot{M} = L_{\rm bol} (GM/R)^{-1}$,
we find the maximum spin up of
\begin{eqnarray}
  \dot{\nu} \leq & 6\times10^{-14}
      \left(\frac{F_{\rm{2.5-25~keV}}}
                 {1\times10^{-9}\rm{~erg~cm}^2\rm{~s}^{-1}}\right)
      \left(\frac{c_{\rm bol}}{2.12}\right)
      \times \nonumber\\ &
      \left(\frac{d}{3.5\rm{~kpc}}\right)^2
      \left(\frac{I}{10^{45}\rm{~g~cm}^2}\right)^{-1}
      \left(\frac{R}{10\rm{~km}}\right)
      \times \nonumber\\ &
      \left(\frac{M}{1.4\ M_\sun}\right)^{-1/3}
      \left(\frac{\nu}{401\rm{~Hz}}\right)^{-1/3}
      \rm{~Hz~s}^{-1}
\end{eqnarray}
Here we use the bolometric correction $c_{\rm bol}$ and distance $d$ from
\citet{Galloway06} and canonical values for the NS parameters.  Integrating
this expression for the observed flux over the 40~d outburst outburst, we can
approximate a physical upper limit for the total spin up of $\Delta\nu \leq
0.12$~\uHz.  For the earlier points to reflect the spin phase, the frequency
shift would need to be $\Delta\nu =
1.1\times10^{-13}$~Hz~s$^{-1}\cdot40\rm{~d} = 0.38$~\uHz.  Unless the source's
luminosity underestimates the accretion rate by a factor of a few, these
points cannot reflect the NS spin phase.

Of course, it is possible that the NS spin phase lies somewhere between these
two extremes.  Based on the data available, we cannot exclude the possibility
of a spin up during the outburst.  Indeed, though the confidence intervals in
1998--2005 include zero, they are biased toward a positive $\dot{\nu}$ (H08),
suggestive that a small sytematic spin up may be present.

\subsection{Interpreting the spin frequency derivative}
\label{sect:DiscSpinDown}

The observed long-term spin down of \saxj\ produces a loss of rotational
energy at a rate of $\dot{E}_{\rm sd} = 4\pi^2 I \nu\dot{\nu} =
-9\times10^{33}$~erg~s$^{-1}$, assuming a canonical value of $I =
10^{45}$~g~cm$^2$ for the NS moment of inertia.  This spin down acts on the NS
during quiescence (H08), and it is consistent with a constant rate over the
ten years of observation.  We consider three possible torques:  magnetic
dipole radiation, the centrifugal expulsion of matter from the inner accretion
disk by the magnetic field (the so-called ``propeller effect''), and
gravitational radiation due to a mass quadrupole moment of the NS.  We assume
that these independent mechanisms contribute additively to the observed
spin-down luminosity, $\dot{E}_{\rm sd} = \dot{E}_{\rm dipole} + \dot{E}_{\rm
prop} + \dot{E}_{\rm gr}$.

The presence of pulsations during the outburst peaks requires $B >
0.4\times10^8$ to truncate the accretion flow above the NS surface
(\citealt{Psaltis99}; H08).  This lower limit implies a magnetic dipole
contribution of $\dot{E}_{\rm dipole} / \dot{E}_{\rm sd} > 10\%$.  If the
field strength is only a few times higher, $B = 1.5\times10^8$~G, then
magnetic dipole radiation would account for the entirety of the long-term spin
down.  These tight limits on $B$ leave a very small window for other
mechanisms.  Their contribution would have to be within a single order of
magnitude of the contribution from magnetic dipole radiation.  Given the wide
physical ranges available to these independent spin-down torques, such a
coincidence is unlikely.  Additionally, the long-term spin down is consistent
with a constant rate, as would be expected if $\dot{E}_{\rm dipole}$
dominates; the constancy of $\dot{E}_{\rm prop}$ and $\dot{E}_{\rm gr}$ is
very model dependent.

Independent estimates of the magnetic field strength suggest that $B$ is
toward the upper end of our allowed interval.  \citet{Gilfanov98} noted that
if the rapid dimming at the end of the main outburst is due to the onset of
the propeller effect (as suggested by other outburst properties, e.g.,
\citealt{Hartman09a}), it would imply $B \sim 1\times10^8$~G.  As noted in the
previous section, the appearance of double-peaked pulse profiles in the tail
suggest a similar field strength \citep{Ibragimov09}.  The detection of a
relativistically broadened iron line during the 2008 outburst also suggests a
higher field of $(3\pm1)\times10^8$~G \citep{Cackett09, Papitto09}.  For all
these reasons, it is likely that magnetic dipole radiation provides the
dominant torque.

\subsection{Interpreting the orbital period derivative}

The cause of the orbital period derivative remains a subject of debate.  H08,
\citet{DiSalvo08}, and \citet{Burderi09} show that it cannot be accounted for
solely by conservative transfer from the low-mass companion to the NS.  We
consider two alternatives: the ejection of mass from the system due to the
ablation of the companion, and the influence of short-term effects that can
exchange angular momentum between the companion and the orbit.

{\em Ejection of matter due to the ablation of the companion.}  If magnetic
dipole torque is the predominant contributor to the spin down of \saxj, as
suggested in the previous section, then the source may behave like a
rotation-powered pulsar during quiescence, producing radio pulsations and a
particle wind.  While no radio pulsations have been detected \citep{Burgay03},
the upper limits are not particularly constraining.  However, during
quiescence the system is substantially brighter in the optical than expected
\citep{Homer01}.  This optical excess may be due to the heating of the
companion by a particle wind \citep{Burderi03}.

If this particle wind ejects a large amount of matter from the system, it
could explain the rapid orbital period change.  For a companion modeled with a
polytropic index of $n = 3/2$ (representative of a degenerate or fully
convective star), a mass loss of $\sim$$10^{-9}\ M_\sun$~yr$^{-1}$ would be
needed to yield the observed $\dot{P}_{\rm orb}$ \citep{DiSalvo08}.

Is the spin-down luminosity of $\dot{E}_{\rm sd} =
-9\times10^{33}$~erg~s$^{-1}$ sufficient to drive such mass loss?  Assuming
that it is radiated isotropically, the incident luminosity onto the companion
is $\dot{E}_{\rm abl} = -\frac{1}{4}(R_c/a)^2\dot{E}_{\rm sd}$, where $R_c$ is
the companion radius and $a$ is the binary separation.  Since the companion
fills its Roche lobe, $R_c/a \approx 0.15$ for a mass ratio of $q \approx
0.05/1.4$ \citep{Eggleton83}.  The resulting luminosity onto the companion is
$\dot{E}_{\rm abl} = 2.1\times10^{32}$~erg~s$^{-1}$.  Assuming perfect
efficiency, that incident luminosity can drive a mass loss of $\dot{M}_c =
-\dot{E}_{\rm abl} R_c / GM_c = -1.2\times10^{-9}\ M_\sun$~yr$^{-1}$.  Thus
this scenario for orbital evolution is energetically feasible.

Nevertheless, some serious problems with this model remain.  A particle wind
sufficient to eject over 95\% of $\dot{M}_c$ from the system, as necessitated
by this model, would also drive away the accretion disk.  \citet{DiSalvo08}
suggest that outbursts therefore occur when the pressure of matter driven from
the companion temporarily exceeds the radiation pressure.  However, the
outburst of \saxj\ and the other AMPs look quite similar to some of the
outbursts form other LMXBs, both in time scales and shape.  The disk
instability model has been generally successful in explaining the light curves
of soft X-ray transients in general \citep[e.g.,][]{King98} and AMPs in
particular \citep{Powell07}.  It is difficult to reconcile these similarities
in light curve shape and time scale with a very different mechanism governing
the outbursts of \saxj.

{\em Short-term changes in the orbital period.}  Alternatively, the observed
$\dot{P}_{\rm orb}$ may not be representative of the secular evolution, but is
instead due to some short-term interchange of angular momentum between the
companion and the orbit.  Similarly large orbital period derivatives are
present in two so-called ``black widow'' millisecond radio pulsars, and these
period derivatives have been observed to change sign on a $\sim$10~yr time
scale \citep{Arzoumanian94, Doroshenko01}.  It has been suggested that tidal
dissipation and magnetic activity in the companion are responsible for the
orbital variability, requiring that the companion is at least partially
non-degenerate, convective, and magnetically active \citep{Arzoumanian94,
Applegate94, Doroshenko01}.  Eclipse timing of the LMXB EXO~0748$-$676 also
shows rapid variation in the orbital period; explanations for the exchange of
angular momentum include a ``wobble'' of the companion's inertial moments due
to differential rotation and convection produced by its uneven heating
\citep{Wolff02}.


However, in all the above systems, the orbital period derivatives change over
$\lesssim$10~yr.  While this does not definitively prove that the
$\dot{P}_{\rm orb}$ of \saxj\ is secular, it would make this source an
exception.  In contrast, the spin-down rate of the ablation model should be
constant, so it is supported by the data from the 2008 outburst.


\bigskip
\acknowledgements{We are grateful to Jean Swank and the \RXTE\ operations team
at NASA Goddard Space Flight Center for their help in scheduling these
observations.  We also thank Tiziana Di Salvo, Luciano Burderi, Duncan
Galloway, and Mike Wolff for useful discussions.  We thank the referee for
useful suggestions.  This work was supported in part by NASA grants NNX07AP93G
and NNX08AJ43G, awarded to MIT through the \RXTE\ Guest Observer Program and
the Astrophysics Data Program.}



\begin{thebibliography}{41}
\expandafter\ifx\csname natexlab\endcsname\relax\def\natexlab#1{#1}\fi

\bibitem[{{Applegate} \& {Shaham}(1994)}]{Applegate94}
{Applegate}, J.~H. \& {Shaham}, J. 1994, \apj, 436, 312

\bibitem[{{Arzoumanian} {et~al.}(1994){Arzoumanian}, {Fruchter}, \&
  {Taylor}}]{Arzoumanian94}
{Arzoumanian}, Z., {Fruchter}, A.~S., \& {Taylor}, J.~H. 1994, \apjl, 426, L85

\bibitem[{{Bildsten} \& {Chakrabarty}(2001)}]{Bildsten01}
{Bildsten}, L. \& {Chakrabarty}, D. 2001, \apj, 557, 292

\bibitem[{{Burderi} {et~al.}(2003){Burderi}, {Di Salvo}, {D'Antona}, {Robba},
  \& {Testa}}]{Burderi03}
{Burderi}, L., {Di Salvo}, T., {D'Antona}, F., {Robba}, N.~R., \& {Testa}, V.
  2003, \aap, 404, L43

\bibitem[{{Burderi} {et~al.}(2006){Burderi}, {Di Salvo}, {Menna}, {Riggio}, \&
  {Papitto}}]{Burderi06}
{Burderi}, L., {Di Salvo}, T., {Menna}, M.~T., {Riggio}, A., \& {Papitto}, A.
  2006, \apjl, 653, L133

\bibitem[{{Burderi} {et~al.}(2009){Burderi}, {Riggio}, {Di Salvo}, {Papitto},
  {Menna}, {D'A{\'i}}, \& {Iaria}}]{Burderi09}
{Burderi}, L., {Riggio}, A., {Di Salvo}, T., {Papitto}, A., {Menna}, M.~T.,
  {D'A{\'i}}, A., \& {Iaria}, R. 2009, \aap, submitted, arXiv: 0902.2128

\bibitem[{{Burgay} {et~al.}(2003){Burgay}, {Burderi}, {Possenti}, {D'Amico},
  {Manchester}, {Lyne}, {Camilo}, \& {Campana}}]{Burgay03}
{Burgay}, M., {Burderi}, L., {Possenti}, A., {D'Amico}, N., {Manchester},
  R.~N., {Lyne}, A.~G., {Camilo}, F., \& {Campana}, S. 2003, \apj, 589, 902

\bibitem[{{Cackett} {et~al.}(2009){Cackett}, {Altamirano}, {Patruno}, {Miller},
  {Reynolds}, {Linares}, \& {Wijnands}}]{Cackett09}
{Cackett}, E.~M., {Altamirano}, D., {Patruno}, A., {Miller}, J.~M., {Reynolds},
  M., {Linares}, M., \& {Wijnands}, R. 2009, \apjl, accepted, arXiv: 0901.3142

\bibitem[{{Campana} {et~al.}(2004){Campana}, {D'Avanzo}, {Casares}, {Covino},
  {Israel}, {Marconi}, {Hynes}, {Charles}, \& {Stella}}]{Campana04}
{Campana}, S., {D'Avanzo}, P., {Casares}, J., {Covino}, S., {Israel}, G.,
  {Marconi}, G., {Hynes}, R., {Charles}, P., \& {Stella}, L. 2004, \apjl, 614,
  L49

\bibitem[{{Campana} {et~al.}(2002){Campana}, {Stella}, {Gastaldello},
  {Mereghetti}, {Colpi}, {Israel}, {Burderi}, {Di Salvo}, \&
  {Robba}}]{Campana02}
{Campana}, S., {Stella}, L., {Gastaldello}, F., {Mereghetti}, S., {Colpi}, M.,
  {Israel}, G.~L., {Burderi}, L., {Di Salvo}, T., \& {Robba}, R.~N. 2002,
  \apjl, 575, L15

\bibitem[{{Chakrabarty} \& {Morgan}(1998)}]{Chakrabarty98}
{Chakrabarty}, D. \& {Morgan}, E.~H. 1998, \nat, 394, 346

\bibitem[{{Chakrabarty} {et~al.}(2003){Chakrabarty}, {Morgan}, {Muno},
  {Galloway}, {Wijnands}, {van der Klis}, \& {Markwardt}}]{Chakrabarty03}
{Chakrabarty}, D., {Morgan}, E.~H., {Muno}, M.~P., {Galloway}, D.~K.,
  {Wijnands}, R., {van der Klis}, M., \& {Markwardt}, C.~B. 2003, \nat, 424, 42

\bibitem[{{Deloye} {et~al.}(2008){Deloye}, {Heinke}, {Taam}, \&
  {Jonker}}]{Deloye08}
{Deloye}, C.~J., {Heinke}, C.~O., {Taam}, R.~E., \& {Jonker}, P.~G. 2008,
  \mnras, 391, 1619

\bibitem[{{Di Salvo} {et~al.}(2008){Di Salvo}, {Burderi}, {Riggio}, {Papitto},
  \& {Menna}}]{DiSalvo08}
{Di Salvo}, T., {Burderi}, L., {Riggio}, A., {Papitto}, A., \& {Menna}, M.~T.
  2008, \mnras, 389, 1851

\bibitem[{{Doroshenko} {et~al.}(2001){Doroshenko}, {L{\"o}hmer}, {Kramer},
  {Jessner}, {Wielebinski}, {Lyne}, \& {Lange}}]{Doroshenko01}
{Doroshenko}, O., {L{\"o}hmer}, O., {Kramer}, M., {Jessner}, A., {Wielebinski},
  R., {Lyne}, A.~G., \& {Lange}, C. 2001, \aap, 379, 579

\bibitem[{{Eggleton}(1983)}]{Eggleton83}
{Eggleton}, P.~P. 1983, \apj, 268, 368

\bibitem[{{Galloway} \& {Cumming}(2006)}]{Galloway06}
{Galloway}, D.~K. \& {Cumming}, A. 2006, \apj, 652, 559

\bibitem[{{Giles} {et~al.}(1999){Giles}, {Hill}, \& {Greenhill}}]{Giles99}
{Giles}, A.~B., {Hill}, K.~M., \& {Greenhill}, J.~G. 1999, \mnras, 304, 47

\bibitem[{{Gilfanov} {et~al.}(1998){Gilfanov}, {Revnivtsev}, {Sunyaev}, \&
  {Churazov}}]{Gilfanov98}
{Gilfanov}, M., {Revnivtsev}, M., {Sunyaev}, R., \& {Churazov}, E. 1998, \aap,
  338, L83

\bibitem[{{Hartman} {et~al.}(2008{\natexlab{a}}){Hartman}, {Patruno},
  {Chakrabarty}, {Kaplan}, {Markwardt}, {Morgan}, {Ray}, {van der Klis}, \&
  {Wijnands}}]{Hartman08}
{Hartman}, J.~M., {Patruno}, A., {Chakrabarty}, D., {Kaplan}, D.~L.,
  {Markwardt}, C.~B., {Morgan}, E.~H., {Ray}, P.~S., {van der Klis}, M., \&
  {Wijnands}, R. 2008{\natexlab{a}}, \apj, 675, 1468 (H08)

\bibitem[{{Hartman} {et~al.}(2008{\natexlab{b}}){Hartman}, {Watts}, \&
  {Chakrabarty}}]{Hartman09a}
{Hartman}, J.~M., {Watts}, A.~L., \& {Chakrabarty}, D. 2008{\natexlab{b}},
  \apj, submitted, arXiv: 0809.3722

\bibitem[{{Heinke} {et~al.}(2007){Heinke}, {Jonker}, {Wijnands}, \&
  {Taam}}]{Heinke07}
{Heinke}, C.~O., {Jonker}, P.~G., {Wijnands}, R., \& {Taam}, R.~E. 2007, \apj,
  660, 1424

\bibitem[{{Homer} {et~al.}(2001){Homer}, {Charles}, {Chakrabarty}, \& {van
  Zyl}}]{Homer01}
{Homer}, L., {Charles}, P.~A., {Chakrabarty}, D., \& {van Zyl}, L. 2001,
  \mnras, 325, 1471

\bibitem[{{Ibragimov} \& {Poutanen}(2009)}]{Ibragimov09}
{Ibragimov}, A. \& {Poutanen}, J. 2009, \mnras, submitted, arXiv: 0901.0073

\bibitem[{{in 't Zand} {et~al.}(2001){in 't Zand}, {Cornelisse}, {Kuulkers},
  {Heise}, {Kuiper}, {Bazzano}, {Cocchi}, {Muller}, {Natalucci}, {Smith}, \&
  {Ubertini}}]{IntZand01}
{in 't Zand}, J.~J.~M., {Cornelisse}, R., {Kuulkers}, E., {Heise}, J.,
  {Kuiper}, L., {Bazzano}, A., {Cocchi}, M., {Muller}, J.~M., {Natalucci}, L.,
  {Smith}, M.~J.~S., \& {Ubertini}, P. 2001, \aap, 372, 916

\bibitem[{{in 't Zand} {et~al.}(1998){in 't Zand}, {Heise}, {Muller},
  {Bazzano}, {Cocchi}, {Natalucci}, \& {Ubertini}}]{IntZand98}
{in 't Zand}, J.~J.~M., {Heise}, J., {Muller}, J.~M., {Bazzano}, A., {Cocchi},
  M., {Natalucci}, L., \& {Ubertini}, P. 1998, \aap, 331, L25

\bibitem[{{King} \& {Ritter}(1998)}]{King98}
{King}, A.~R. \& {Ritter}, H. 1998, \mnras, 293, L42

\bibitem[{{Long} {et~al.}(2008){Long}, {Romanova}, \& {Lovelace}}]{Long08}
{Long}, M., {Romanova}, M.~M., \& {Lovelace}, R.~V.~E. 2008, \mnras, 386, 1274

\bibitem[{{Markwardt} \& {Swank}(2008)}]{Markwardt08}
{Markwardt}, C.~B. \& {Swank}, J.~H. 2008, The Astronomer's Telegram, 1728

\bibitem[{{Papitto} {et~al.}(2009){Papitto}, {di Salvo}, {D'A{\i}}, {Iaria},
  {Burderi}, {Riggio}, {Menna}, \& {Robba}}]{Papitto09}
{Papitto}, A., {di Salvo}, T., {D'A{\i}}, A., {Iaria}, R., {Burderi}, L.,
  {Riggio}, A., {Menna}, M.~T., \& {Robba}, N.~R. 2009, \aap, 493, L39

\bibitem[{{Patruno} {et~al.}(2008){Patruno}, {Hartman}, {Wijnands}, {van der
  Klis}, {Chakrabarty}, {Morgan}, \& {Markwardt}}]{Patruno08}
{Patruno}, A., {Hartman}, J.~M., {Wijnands}, R., {van der Klis}, M.,
  {Chakrabarty}, D., {Morgan}, E.~H., \& {Markwardt}, C.~B. 2008, The
  Astronomer's Telegram, 1760

\bibitem[{{Powell} {et~al.}(2007){Powell}, {Haswell}, \& {Falanga}}]{Powell07}
{Powell}, C., {Haswell}, C., \& {Falanga}, M. 2007, \mnras, 374, 466

\bibitem[{{Pringle} \& {Rees}(1972)}]{Pringle72}
{Pringle}, J.~E. \& {Rees}, M.~J. 1972, \aap, 21, 1

\bibitem[{{Psaltis} \& {Chakrabarty}(1999)}]{Psaltis99}
{Psaltis}, D. \& {Chakrabarty}, D. 1999, \apj, 521, 332

\bibitem[{{Roche} {et~al.}(1998){Roche}, {Chakrabarty}, {Morales-Rueda},
  {Hynes}, {Slivan}, {Simpson}, \& {Hewett}}]{Roche98}
{Roche}, P., {Chakrabarty}, D., {Morales-Rueda}, L., {Hynes}, R., {Slivan},
  S.~M., {Simpson}, C., \& {Hewett}, P. 1998, \iaucirc, 6885

\bibitem[{{Wang} {et~al.}(2008){Wang}, {Bassa}, {Cumming}, \& {Kaspi}}]{Wang08}
{Wang}, Z., {Bassa}, C., {Cumming}, A., \& {Kaspi}, V.~M. 2008, \apj, accepted,
  arXiv: 0812.2815

\bibitem[{{Wang} {et~al.}(2001){Wang}, {Chakrabarty}, {Roche}, {Charles},
  {Kuulkers}, {Shahbaz}, {Simpson}, {Forbes}, \& {Helsdon}}]{Wang01}
{Wang}, Z., {Chakrabarty}, D., {Roche}, P., {Charles}, P.~A., {Kuulkers}, E.,
  {Shahbaz}, T., {Simpson}, C., {Forbes}, D.~A., \& {Helsdon}, S.~F. 2001,
  \apjl, 563, L61

\bibitem[{{Wijnands} {et~al.}(2001){Wijnands}, {M{\'e}ndez}, {Markwardt}, {van
  der Klis}, {Chakrabarty}, \& {Morgan}}]{Wijnands01}
{Wijnands}, R., {M{\'e}ndez}, M., {Markwardt}, C., {van der Klis}, M.,
  {Chakrabarty}, D., \& {Morgan}, E. 2001, \apj, 560, 892

\bibitem[{{Wijnands} \& {van der Klis}(1998)}]{Wijnands98}
{Wijnands}, R. \& {van der Klis}, M. 1998, \nat, 394, 344

\bibitem[{{Wijnands} {et~al.}(2003){Wijnands}, {van der Klis}, {Homan},
  {Chakrabarty}, {Markwardt}, \& {Morgan}}]{Wijnands03}
{Wijnands}, R., {van der Klis}, M., {Homan}, J., {Chakrabarty}, D.,
  {Markwardt}, C.~B., \& {Morgan}, E.~H. 2003, \nat, 424, 44

\bibitem[{{Wolff} {et~al.}(2002){Wolff}, {Hertz}, {Wood}, {Ray}, \&
  {Bandyopadhyay}}]{Wolff02}
{Wolff}, M.~T., {Hertz}, P., {Wood}, K.~S., {Ray}, P.~S., \& {Bandyopadhyay},
  R.~M. 2002, \apj, 575, 384

\end{thebibliography}
\end{document}